\documentclass[aps, prb, twocolumn, showpacs, floatfix]{revtex4-1}
\usepackage{amsmath, amsfonts, amssymb, mathbbol}

\usepackage{graphicx,color}
\usepackage{hyperref}

\begin{document}

\title{Graphene quantum dot on boron nitride: Dirac cone replica and
  Hofstadter butterfly}

\newcommand{\itptuw}{}

\author{L.A.~Chizhova}
\affiliation{Institute for Theoretical Physics, Vienna
  University of Technology, Wiedner Hauptstra{\ss}e 8-10, 1040 Vienna,
  Austria, EU}

\author{F.~Libisch}
\email{florian.libisch@tuwien.ac.at}
\affiliation{Institute for Theoretical Physics, Vienna
  University of Technology, Wiedner Hauptstra{\ss}e 8-10, 1040 Vienna,
  Austria, EU}

\author{J.~Burgd\"orfer}
\affiliation{Institute for Theoretical Physics, Vienna
  University of Technology, Wiedner Hauptstra{\ss}e 8-10, 1040 Vienna,
  Austria, EU}

\date{\today}

\begin{abstract}
Graphene flakes placed on hexagonal boron nitride feature in the
presence of a magnetic field a complex electronic structure due to a
hexagonal moir\'e potential resulting from the van der Waals
interaction with the substrate. The slight lattice mismatch gives rise
to a periodic supercell potential. Zone folding is expected to create
replica of the original Dirac cone and Hofstadter butterflies. Our
large-scale tight binding simulation reveals an unexpected coexistence
of a relativistic and non-relativistic Landau level structure. The
presence of the zeroth Landau level and its associated butterfly is
shown to be the unambiguous signature for the occurrence of Dirac cone
replica.
\end{abstract}


\maketitle

\section{Introduction}
\label{sec:section1}
The electronic and transport properties of graphene strongly depend on
the substrate graphene is placed on or substituted in. For instance,
in the case of widely-used SiO$_2$, the roughness of the substrate
surface introduces a corrugation of the graphene
monolayer,\cite{Geringer09} puddles,\cite{Martin07} and charge
traps.\cite{Neto09} Reducing substrate-induced disorder is critical
for achieving higher carrier mobility especially in transport
applications. Graphene on clean transition metal surfaces
[e.g., iridium (Ir) \cite{Voloshina13} or ruthenium \cite{Marchini07}]
or on graphite \cite{GuohongLi09} have been shown to feature much reduced
disorder.\cite{Dean10, Pletikosic09} More recently, the wide gap
insulator hBN received major attention as substrate material since it
is inert to the carriers in graphene near the Fermi energy.
\cite{Xue11, Decker11}

A small lattice mismatch between graphene and the substrate leads to
periodic potential modulations on a scale much larger than the lattice
vector.\cite{Park08, Yankowitz12, Subramaniam12} Such a potential is
formed, for instance, by a grid of electron-beam deposited adatoms on
graphene,\cite{Park08, Meyer08} by the misalignment of graphene layers
in twisted bilayer graphene,\cite{Bistritzer11, Moon13} or by a small
lattice mismatch between graphene and a hexagonal substrate (BN or
Ir), resulting in a so-called moir\'e pattern.\cite{Xue11, Decker11}
For hexagonal boron nitride, the layer-substrate interaction is of Van
der Waals type. The effect of the substrate can be, to first order,
captured by an additional smooth periodic potential with superlattice
periodicity $a^S$ large compared to the lattice periodicity $a$. For
graphene on hBN and an alignment angle $\phi=0^\circ$, $a^S = 13.8$ nm
($a^S/a \gtrsim 50$). Introducing the additional length scale $a^S$
into the physics of graphene devices gives rise to interesting new
phenomena. The zone folding due to the periodic superlattice leads to
additional high symmetry points in the bandstructure.\cite{Park08} DFT
simulations of a supercell of graphene on hBN and STM measurements
suggest additional Dirac cones in the bandstructure.\cite{Yankowitz12}
Indeed, recent experiments of magnetotransport \cite{Gorbachev14} and
quantum capacitance effects in a magnetic field \cite{Hunt13} observe
the formation of replica of Landau level structures energetically
above and below the primary structure associated with the ``main''
Dirac point. These structures were attributed\cite{Gorbachev14} to
satellite Dirac-cones caused by the moir\'e pattern of the
superlattice. However, a zeroth Landau level at the satellite, the
hallmark of Dirac-like Landau level structures \cite{Rabi28} is
conspicuously missing.\cite{Gorbachev14,Hunt13} In the present work,
we aim at elucidating the origin of these satellites, and explaining
why experiments have, up to now, failed to reproduce the expected
zeroth Landau level in the satellite structures.

\begin{figure}[t]

    \includegraphics[width=\linewidth]{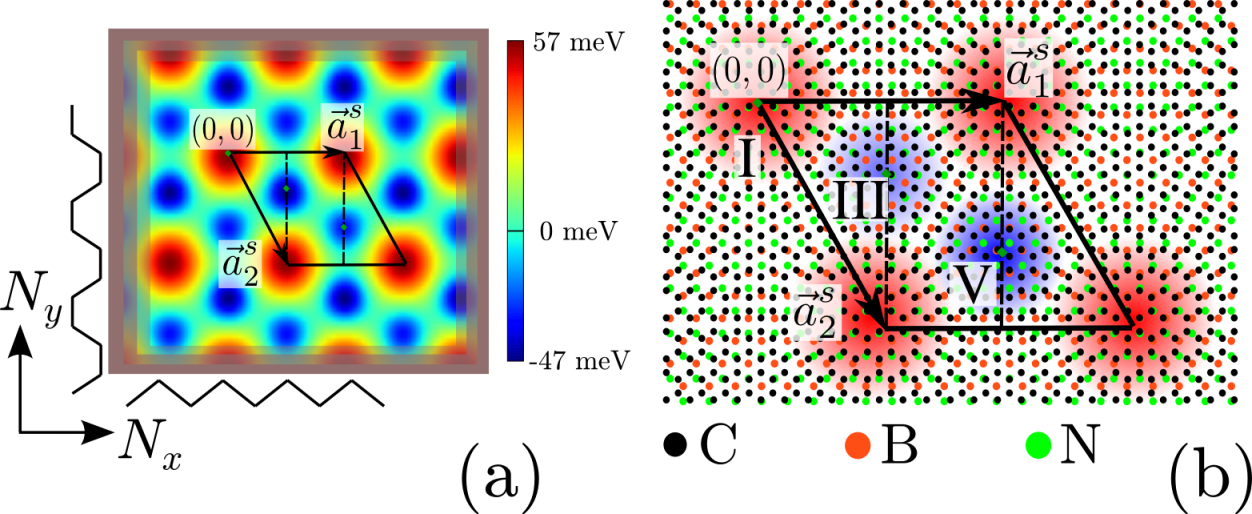}

    \caption{(a) Schematic view of graphene flake with the superlattice
      potential landscape, (b) superlattice unit cell for graphene on
      BN with carbon (black), boron (orange) and nitrogen (green) atoms.
      The areas $I$ (red), $I\!I\!I$ and $V$ (blue)
      are taken from Ref.~\onlinecite{Katsnelson11}}

\label{fig:dot}
\end{figure}

Based on recent ab-initio DFT calculations,\cite{Katsnelson11,Roche14}
we simulate a realistic, extended graphene nanoflake interacting with
an hBN substrate (Fig.~\ref{fig:dot}). We benchmark our description by
reproducing characteristic features of graphene on hBN in a magnetic
field, e.g., the Hofstadter butterfly,\cite{Koshino06, Hunt13, Dean13,
  Hofstadter76} as well as the observed satellite structures.  Our
results suggest that these satellites are caused by parabolic extremal
points in the bandstructure giving rise to Schr\"odinger-like Landau
levels rather than replica of Dirac cones. Key is the observation
that, coincidentally, both Dirac- and Schr\"odinger like dispersion
relations give rise to linear Landau level structures when plotted as
a function of the back gate voltage. The distinguishing feature turns
out to be the presence (or absence) of a magnetic-field independent
zeroth Landau level. Only when employing an unrealistically strong
superlattice potential true Dirac-cone like satellite structures, that
include a zeroth Landau level, emerge close to the main Dirac point.

\section{Graphene model Hamiltonian}
\label{sec:section2}

While our numerical simulation employs a third-order tight-binding
Hamiltonian (for details see Ref.~\onlinecite{Libisch10})
realistically reproducing the bandstructure of graphene, it is
instructive for the discussion of superlattice effects to briefly
recall the essential features of the continuous low-energy
approximation in terms of an effective massless Dirac Hamiltonian with
the spinor components related to the sublattice degree of freedom,
($\psi_A,\psi_B$), the so-called pseudospin.\cite{Neto09, Sarma11}
Including the physical spin, one arrives at a four-component spinor.\cite{Libisch10} In the
presence of a homogeneous magnetic field $\vec{B}=\vec{\nabla} \times
\vec{A}$ perpendicular to the graphene plane as well as an external
electrostatic potential the Dirac Hamiltonian reads
\begin{align}
\label{eq:Dirac_pot}
H_D =\ & v_F \vec{\sigma} \cdot ( \hat{\vec{p}} - \frac{e}{c}\vec{A} )
\otimes \tau_1 + v_F \vec{\sigma^*} \cdot ( \hat{\vec{p}} -
\frac{e}{c}\vec{A} ) \otimes \tau_2 \nonumber \\ & + V(\vec{r})\cdot
\sigma_0 \otimes \tau_0 + W(\vec{r})\cdot \sigma_z\otimes\tau_0
\end{align}
where the Pauli matrices $\sigma_{x,y,z}$ ($\tau_{x,y,z}$) and unit
matrix $\sigma_0$ ($\tau_0$) act on the $A$-$B$ sublattice ($K$-$K'$
or valley) degree of freedom, $\tau_{1,2} = (\tau_0 \pm \tau_z)/2$ ,
$e$ is the elementary charge and $c$ is the speed of light.  In
Eq.~(\ref{eq:Dirac_pot}) we have distinguished two different classes
of potentials: the (slowly varying) background potential $V(\vec{r})$
(represented by the unit matrix in sublattice space) breaking the
particle-hole symmetry within the Dirac double cone and the (short
range) contribution $W(\vec{r})$ (represented by $\sigma_z$ in
sublattice space) breaking sublattice symmetry.  $W(\vec{r})$
effectively introduces a finite ``mass'' for the Dirac fermions and
thus a band gap at the Dirac point.\cite{Berry87}

The effect of the hBN substrate can now be modeled by judicious
choices for the potentials $V(\vec r)$ and $W(\vec r)$.  The moir\'e
pattern of graphene on hBN with an alignment angle of $\phi \approx
0^{\circ}$ features a period of $a^S=13.8$ nm and hexagonal
superlattice vectors $\vec{a}^S_1 = (13.8, 0) \mbox{ nm}$ and
$\vec{a}_2^S = (6.9, 11.9) \mbox{ nm}$ [see
  Fig.~\ref{fig:dot}(b)]. The supercell may be partitioned into five
regions based on the relative local alignment of the graphene
and hBN layer [see labels in Fig.~\ref{fig:dot}(b) and
  Ref.~\onlinecite{Katsnelson11}]: in region $I$, the carbon atoms of
one sublattice $A$ are on top of boron and the carbon atoms of the
other sublattice $B$ on top of nitrogen; the region $I\!I\!I$ features
the carbon atoms of $A$ on top of the nitrogen atoms and the atoms of
$B$ are in the middle off a BN hexagon while in region $V$ the carbon
atoms of $A$ are on top of boron atoms while now the $B$ atoms are
located off the BN hexagons.
We deduce realistic potential parameters
from recent ab-initio DFT calculations,\cite{Katsnelson11, Roche14}
where sublattice symmetry breaking potentials of type $W(\vec{r})$
feature broad maxima and minima at the centers of the regions $I$, $I\!I\!I$
and $V$, while the transition regions $I\!I$ and $I\!V$ feature intermediate
stacking configurations and potential values.  We thus expand
$W(\vec{r})$ in Gaussians according to
\begin{equation}\label{eq:superlat_pot}
W(\vec{r}) =\sum_{i=I,I\!I\!I,V} W_i \exp \left(-\frac{(\vec{r}-\vec{R}_{i})^2}{2w_i^2}\right)
\end{equation}
with amplitudes $W_I = 57$ meV, $W_{I\!I\!I} = -34$ meV, $W_{V} = -47$
meV taken from Sachs et al.,\cite{Katsnelson11} and widths $0.63 \cdot
w_{I}=w_{I\!I\!I}=w_{V}=7$ nm from geometrical considerations. We note
that local doping by, e.g., charge traps may lead to further local
potential variation in the experiment. While the potential $W(\vec r)$
opens a gap near the Dirac point, the substrate interaction
represented by the potential $V(\vec r)$ breaks the electron-hole
symmetry of the Dirac Hamiltonian (Eq.~\ref{eq:Dirac_pot}). Note that
in the numerical solution employing a third-order TB Hamiltonian (see
below) the exact particle-hole symmetry is already weakly broken in
the absence of $V$. The experimental data, indeed, reveals a
pronounced asymmetry between the electron and hole
satellites.\cite{Gorbachev14} DFT calculations of the adhesion energy
of graphene on hBN suggest only 20 meV stronger binding in region V
than in other regions.\cite{Katsnelson11} As we have verified
numerically, this estimate is too small to reproduce the
experimentally observed asymmetry. Yankowitz et al.\cite{Yankowitz12}
estimate the variation of $V(\vec r)$ from second-order perturbation
theory to be of the order of $120$ meV. We expand $V(\mathbf{r})$ in
terms of Gaussians of the form of Eq.~(\ref{eq:superlat_pot}), with
the amplitudes of $V_V \approx -100$ meV, $V_{I}=V_{I\!I\!I}=0$,
placing the potential minimum at the site of the strongest adhesion.

The experiment indicates the complete lifting of the four-fold
degeneracy of the zeroth Landau level. In addition to the Zeeman
splitting of the spin degree of freedom with $\Delta
E_Z=g_{\mathrm{eff}}\mu_B B$ also the valley degeneracy is lifted by
exchange interaction related to the energy cost of a spin reversal
relative to adjacent (polarized) spins.\cite{Young12} Measurements of
quantum Hall states as a function of magnetic field suggest a linear
increase of valley splitting with magnetic field, which is, to our
knowledge, currently not fully understood theoretically.\cite{Young12}
To account for such a many-body (MB) effect within our single-particle
description, we add to the potential $W(\vec{r})$ a phenomenological
correction
\begin{align}
\label{eq:mp_pot}
W_{\mathrm{MB}}(\vec r) = \alpha \cdot B e^{-r^2/2w_I^2},
\end{align}
that scales linearly in $B$ with $\alpha = 8$ meV/T taken from
experiment.\cite{Young12} The spin coupling also enhances the
gyromagnetic ratio $g_{\mathrm{eff}}$ governing the Zeeman effect
relative to its bare value $g_0=2$.

\begin{figure*}[tb]
    \centering
    \includegraphics[width=\textwidth]{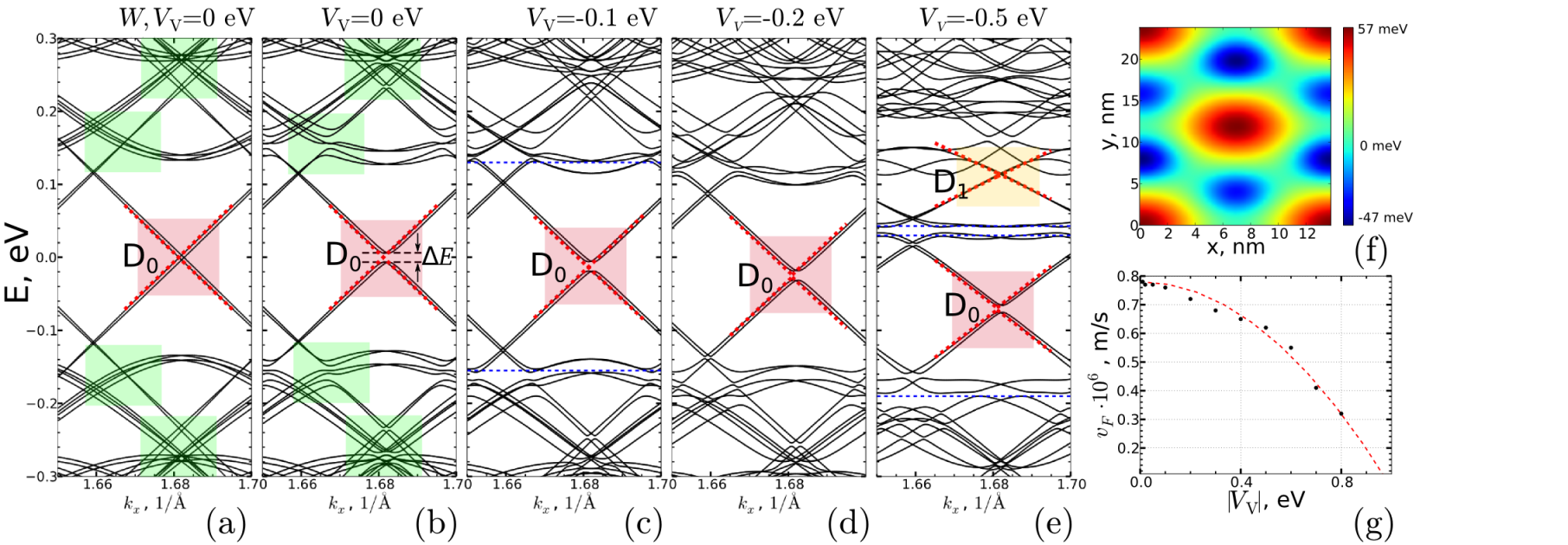}

    \caption{Cut through bandstructure ($k_y=0$) of bulk graphene with
      periodic superlattice potential within the reduced zone
      structure of the superlattice calculated using third
      nearest-neighbor tight-binding model: (a) $V=W=0$ eV,
      i.e.~free-standing graphene, (b)-(e) $W_I = 57$ meV, $W_{I\!I\!I} =
      -34$ meV and $W_V = -47$ meV with varying $V_V$.  The
      red dashed lines show a linear fit to the main Dirac cone D$_0$
      (red regions) and to the secondary cone D$_1$ (yellow area) in
      (e), from which values of the Fermi velocity $v_F$ are
      extracted. Green regions in (b) shows the formation of avoided
      crossings due to the moire potential, which are absent in
      (a). (f) Potential landscape of the supercell used in the
      bandstructure calculations.  (g) The dependence of the Fermi
      velocity $v_F$ of D$_0$ on the amplitude of the superlattice
      potential $V_V$.}

\label{fig:bandstruct}
\end{figure*}

To elucidate the evolution of the spectrum of a graphene flake in the
presence of a superlattice potential, we first consider ideal,
free-standing graphene in the Dirac approximation
[Eq.~(\ref{eq:Dirac_pot})], with $V= W =0$. In the presence of a
perpendicular magnetic field one obtains the Landau level spectrum for
Dirac fermions\cite{Rabi28}
\begin{equation}
\label{eq:landaulevels}
E^D_n(B) = \text{sgn}(n)\sqrt{2|e|\hbar v_F^2|n|B},\ \ \ n\in Z_0,
\end{equation}
This spectrum has three prominent features: (i) the existence of a
0$^{\mathrm{th}}$ Landau level ($n=0$), which does not depend on the
magnetic field, (ii) a graphene-specific degeneracy of the Landau
levels due to the valley symmetry. and (iii) a square root dependence
of all $n\ne 0$ Landau levels on the magnetic field. This
non-equidistant spacing provides a clear-cut distinction to the
equidistant level spacing of non-relativistic Schr\"odinger electrons
where the Landau level spectrum takes on the form of a harmonic
oscillator,\cite{Landau:QM} $E^S_n(B) = \hbar \omega_B(n+1/2)$.

\begin{figure*}[tb]
    \centering
    \includegraphics[width=0.93\textwidth]{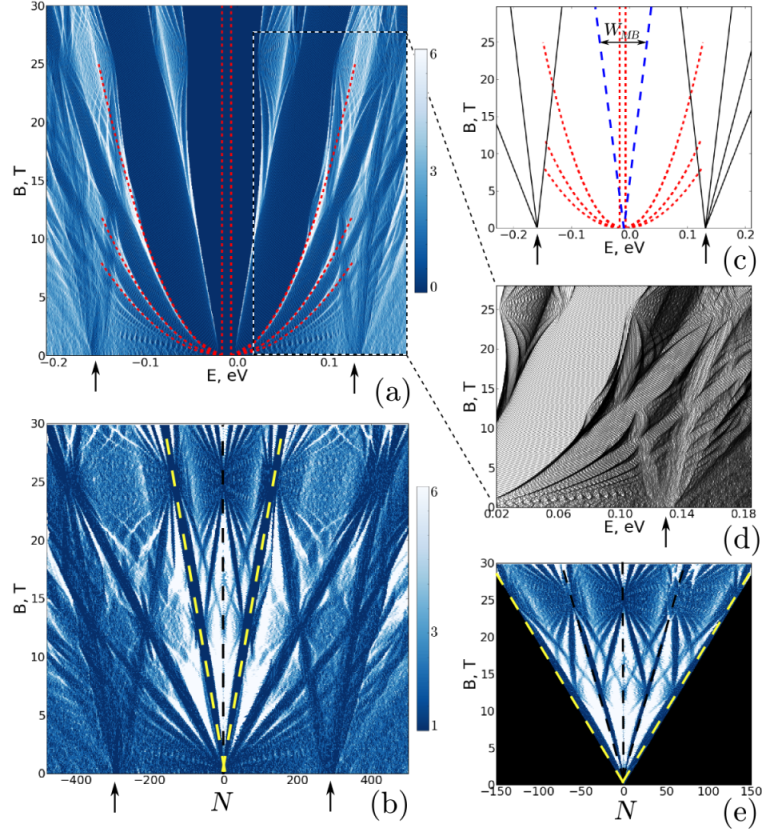}
     \caption{Density of states of a graphene flake in the presence of
       a superlattice potential ($V_V=-0.1$ eV): (a) as a function of
       electron energy and magnetic field; (b) same as (a) however as
       a function of the number of charge carriers $N$; (c) schematic
       plot of the important Landau level structures seen in (a): red
       dashed lines represent the Landau levels of the main Dirac cone
       including the energy gap $\Delta E=10$ meV [see
         Fig.~\ref{fig:bandstruct}]; blue dashed lines denote the
       splitting of the zeroth Landau level due to $W_{\mathrm{MB}}$;
       solid lines represent the Landau levels of the two
       Schr\"odinger-like satellites, whose origin is marked with
       black arrows in (a)-(d).  (d) magnification of the area with
       the right satellite structure in (a).  (e) Four-fold splitting
       of the 0$^{\mathrm{th}}$ Landau level including
       $W_{\mathrm{MB}}$ and Zeeman term with an enhanced $g=5$ due to
       electron-electron interaction [compare the same parameter
         region confined by yellow dashed lines in (b), where the
         Zeeman term is not included].}

\label{fig:DOS_weak}
\end{figure*}

In current experiments, however, the relation between magnetic field
and the back gate voltage ($V_{\mathrm{BG}}$) rather than the energy
$E$ is accessible.  It is therefore instructive to inquire into the
mapping of Dirac-like and Schr\"odinger-like Landau levels onto the
$V_{\mathrm{BG}}$-$B$ plane. Applying a back-gate voltage
$V_{\mathrm{BG}}$ induces a charge $Q$ (proportional to
$V_{\mathrm{BG}}$) on the graphene flake by capacitive coupling. This
in turn changes the Fermi level of the graphene flake.  A capacitive
coupling model predicts, due to the linear density of states (DOS) of
Dirac electrons, a square-root relation\cite{Reiter14} between
$V_{\mathrm{BG}}$ and $E$, $E \propto
\sqrt{V_{\mathrm{BG}}}$. Therefore, Eq.~(\ref{eq:landaulevels})
predicts a linear $V_{\mathrm{BG}}$-$B$ relation. Remarkably, the same
holds for the non-relativistic Schr\"odinger electrons since the
nonrelativistic density of states in 2$D$ is energy independent,
$\rho_S=$const, and $Q\sim V_{\mathrm{BG}}\sim E$. Consequently, the
linear relation between $V_{\mathrm{BG}}$ and $B$ or, equivalently,
$V_{\mathrm{BG}}$ and the charge carrier number $N$ is found for both
a Dirac-like and Schr\"odinger-like spectrum and cannot be used to
reliably identify a Dirac cone or its replica. Instead, the
distinctive feature is therefore the presence or absence of the $n=0$
level.

Another closely related and remarkable feature is the Hofstadter
butterfly,\cite{Koshino06, Hunt13, Dean13, Hofstadter76} observed in
the simultaneous presence of the spatially periodic superlattice and
the $2\pi$ periodic magnetic phase characterized by the magnetic flux
through the area $A$ of one unit cell of the
superlattice,\cite{Koshino06, Geisler04, Hofstadter76}
$\phi / \phi_0 = BAe/h$.
Because of
the large lattice constant of the supercell, the Hofstadter butterfly
becomes accessible at moderate laboratory field strength of the $B$
field.

\begin{figure}[t]

    \centering
    \includegraphics[width=.45\linewidth]{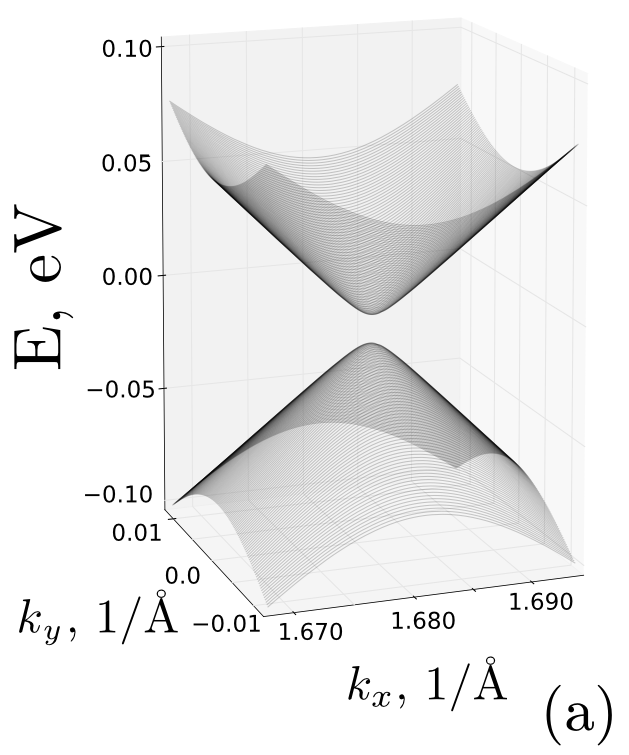}
    \includegraphics[width=.45\linewidth]{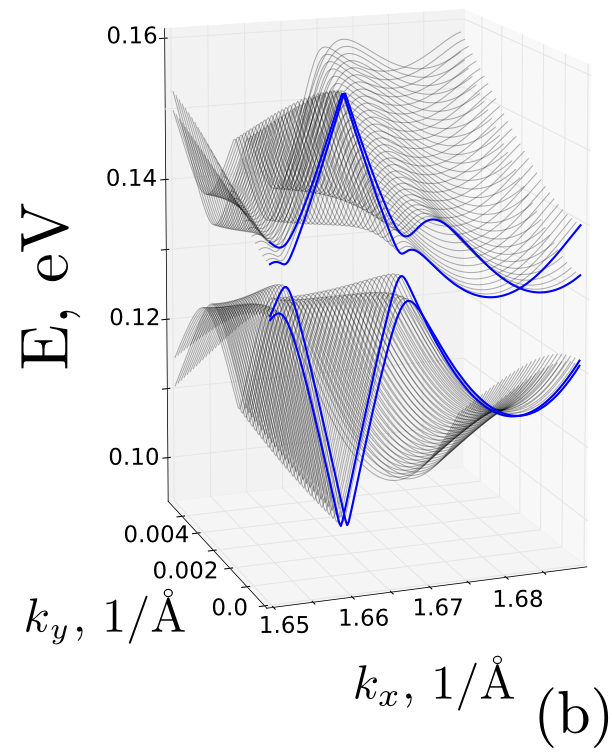}

    \caption{
     Two-dimensional zero-field ($B=0$) bandstructure $E(k_x)$. (a) Near
     the main Dirac cone D$_0$ (see Fig.\ref{fig:bandstruct}(c)).
     (b) Bandstructure in the region of the satellite structure marked by the
     right arrow in Fig.\ref{fig:DOS_weak}(a) and (c).
    }
\label{fig:3D_weak}
\end{figure}

\begin{figure}[t]

    \centering
    \includegraphics[width=.45\linewidth]{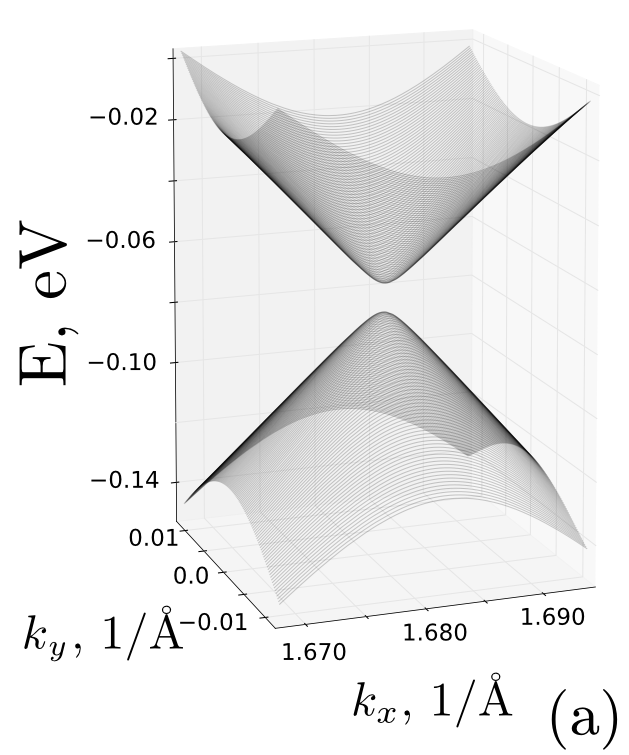}
    \includegraphics[width=.45\linewidth]{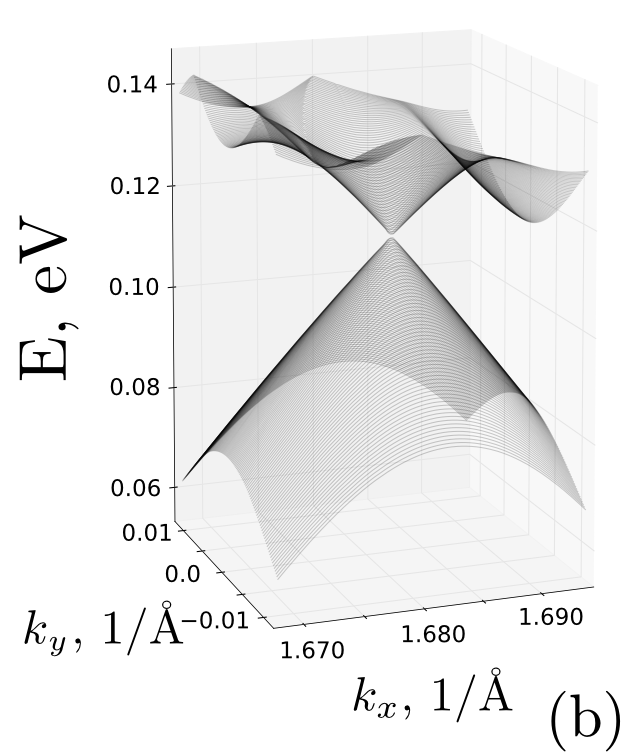}

    \caption{ Two-dimensional band structure for graphene on hBN with an unrealistically
    enhanced superlattice potential ($V_V=-0.5$ eV).
    (a) Primary Dirac cone D$_0$, (b) secondary Dirac cone D$_1$ (see Fig.\ref{fig:bandstruct}(e))}
\label{fig:3D_strong}
\end{figure}

\begin{figure*}[tb]

    \centering
    \includegraphics[width=1.0\textwidth]{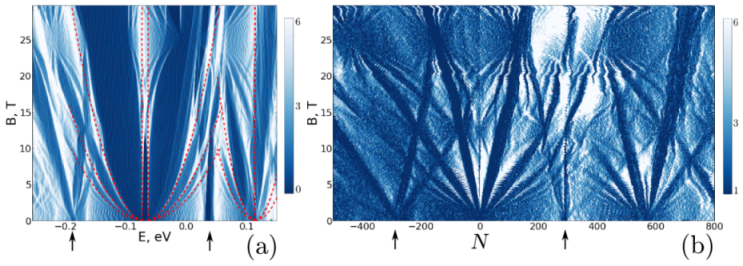}

    \caption{Density of states of a graphene flake on hBN however with
      an unrealistically large superlattice potential ($V_V=-0.5$ eV
      and $W_{\mathrm{MB}}=0$): (a) in the $B$-$E$ plane, as a
      function of magnetic field and electron energy; (b) in the $B$-$N$
      plane, as a function of magnetic field and number of charge
      carriers. Black arrows at $B=0$ mark the same Schr\"odinger
     like satellites as in Fig.~\ref{fig:DOS_weak}.}

\label{fig:DOS_strong}
\end{figure*}

\section{Simulations}
\label{sec:section3}

We simulate the response of a finite-sized patch of graphene with
dimensions $N_x \times N_y = 170 \times 140 \mbox{ nm}^2$ (see
Fig.~\ref{fig:dot}(a)), surrounded by a repulsive edge potential to
eliminate edge effects, using a third-nearest neighbor tight-binding
approach as outlined in Ref.~\onlinecite{Libisch10}. We include the
hBN substrate and many-particle effects through the potentials of
Eqs.(\ref{eq:superlat_pot})-(\ref{eq:mp_pot}) discussed above, the
magnetic field by a Peierls phase factor\cite{Peierls}, and account
for the Zeeman term by first-order perturbation theory.


We first consider the unperturbed [Fig.~\ref{fig:bandstruct}(a)] and
the superlattice-induced bandstructure
[Fig.~\ref{fig:bandstruct}(b)-(e)]. To calculate the bandstructure we use
a supercell with the potential distribution depicted in Fig.~\ref{fig:bandstruct}(f)
imposing periodic boundary conditions.
The unperturbed band structure ($V=W=0$) is displayed within the first
Brillioun zone of the reciprocal supercell. The parabolic bands
[Fig. 2(a)] correspond to cuts through back-folded off-center ($k_y\ne
0$) cones while the near linear bands correspond to the main cone
D$_0$ and its replica centered at $k_y=0$. In the presence of the
substrate interactions $W(\vec{r})$, the band gap $\Delta E$ for the main Dirac
cone is $\Delta E=10$ meV, consistent with the
experiment.\cite{Hunt13} Moreover, the presence of $W$ induces
numerous avoided crossings [green areas in
  Fig.~\ref{fig:bandstruct}(b)] which are absent in free-standing
graphene [green areas in Fig.~\ref{fig:bandstruct}(a)].  Adding the
on-site potential $V(\mathbf r)$ further enhances the particle-hole
asymmetry (beyond the small assymetry of the graphene bandstructure
as captured by the third nearest-nearest neighbor approximation \cite{Reich02}),
shifts the position of the main cone D$_0$ relative to $E=0$ eV by (to
first order) the average of the superlattice potential
[Fig.~\ref{fig:bandstruct}(c)-(e) red area] and enhances the size of
the avoided crossings. Furthermore, the effective Fermi velocity $v_F$
of the main Dirac cone [Fig.~\ref{fig:bandstruct}(a)-(e) red area]
decreases with increasing amplitude $V_V$
[Fig.~\ref{fig:bandstruct}(g)]. In the case of free-standing graphene
the Fermi velocity is $v_{F0} = 0.78 \cdot 10^6$ m/s. With increasing
$|V_V|$ the velocity $v_F$ drops proportionally to the square of the
amplitude of the on-site potential $V_V$, in agreement with
second-order perturbation theory,\cite{Park08b} which predicts a
velocity renormalization
\begin{equation}
v_{F\vec{k}} = v_{F0} - \sum \limits_{\vec{G} \neq 0} \frac{2 \left| V(\vec{G})\right|^2}{v_{F0} \left| \vec{G} \right|} \sin^2{\theta_{\vec{k}, \vec{G}}}.
\end{equation}
$\vec{G}$ is the reciprocal lattice vector, $V(\vec{G})$ is the
Fourier transform of the superlattice potential $V(\vec{r})$ of the form of
Eq.(\ref{eq:superlat_pot}), and $\theta_{\vec{k}, \vec{G}}$ is the
angle between $\vec{k}$ and $\vec{G}$. For a finite alignment 
angle $\phi$ between the hBN layer and the graphene, the periodicity
$a^S$ of the moir\'e pattern is decreased, thereby increasing the size
of the Billouin zone in reciprocal space. Consequently, the 
additional bands due to the overlap between the main Dirac cone and
its replica lie further away from the Dirac point, as we have
verified numerically.

We present the DOS for a realistic value of $V_V=-0.1$ eV in both the
$E$-$B$ plane [Fig.~\ref{fig:DOS_weak}(a,d)] and the $N$-$B$ plane
[Fig.~\ref{fig:DOS_weak}(b)] where $N$ is the number of charge
carriers in the quantum dot ($\propto V_{\mathrm{BG}}$ within a linear
capacitance model). For the transformation from energy to charge
carrier number we do not explicitly use the square root dependence
$E\sim\sqrt N$ for bulk graphene but sum over the number of
eigenstates (or charge) of the finite size flake lying between the
Dirac cone and the appropriate Fermi energy $E_F$,
to accurately account for deviations
from the linear DOS close to the Dirac point.

The calculated DOS displays the formation of Landau levels emanating
from the Dirac point of the main cone D$_0$
[Fig.~\ref{fig:DOS_weak}(a,c)]. The parabolic dependence
[Eq.~(\ref{eq:landaulevels})] $B\sim E^2$ [see red dashed parabolas
  in Fig.~\ref{fig:DOS_weak}(a) and the schematic plot in (c)]
confirms a relativistic diamagnetic response behavior. Moreover, the
curvature of the parabolas determined by the Fermi velocity
$v_{F}=0.76 \cdot 10^6$ m/s [Eq.~(\ref{eq:landaulevels})] extracted
from the fit to the main Dirac cone in the bandstructure [red dashed
  lines in Fig.\ref{fig:bandstruct}(c)] agrees well with that of the
simulated DOS. The zeroth Landau level splits linearly due to the
many-body correction [Eq.~(\ref{eq:mp_pot})]. As discussed in Sec.~II,
the Landau levels as a function of $V_{\mathrm{BG}}$ (or here as a
function of charge carrier number $N$) increase linearly with the
magnetic field $B$ [see Fig.~\ref{fig:DOS_weak}(b)] in agreement with
experiment [see, for example, Refs.~\onlinecite{Gorbachev14,
    Hunt13}]. Our numerical data also reproduce the pronounced
splitting of the four-fold degeneracy of the zeroth Landau level
[Fig.~\ref{fig:DOS_weak}(e)] when including the B-field dependent
many-body term $W_{\mathrm{MB}}$ and the Zeeman term with strongly
enhanced gyromagnetic ratio\cite{Young12} $g_{\mathrm{eff}}=5$. Note
the effective $g$ factor will, in general, be different for different
Landau levels. Our effective many-body potential $W_{\mathrm{MB}}$ has
negligible influence on energies far away from the Dirac point, e.g., on the
satellite structures.  It should be noted that our simulation does not
reproduce a gap at the Dirac point as function of $N$ seen in the
experiment [see Fig.~\ref{fig:DOS_weak}(c)]. The latter results from
quantum capacitance effects \cite{Reiter14} not included in our
simulation.

Superlattice-induced effects on the diamagnetic spectrum are
prominently visible: we observe a Hofstadter spectrum with
"diamond"-like structures \cite{Koshino06} which are most pronounced
at rational fractions\cite{Koshino06, Geisler04} of $\phi/\phi_0$. For
the hexagonal superlattice with a period of $a^{S}=13.8$ nm, this
ratio equals one at $B_0=25.5$ T [see Fig.~\ref{fig:DOS_weak}(a,
  d)]. Moreover, we observe two distinct satellite features which
evolve (approximately) linearly in the $E$-$B$ plane emerging at $B=0$
near $E=0.13$ eV and $-0.16$ eV [marked by arrows in
  Fig.~\ref{fig:DOS_weak}(a, c, d)]. These satellite structures
clearly display a non-relativistic rather than a relativistic $E(B)$
dependence. A closer look into the $B=0$ bandstructure
[Fig.~\ref{fig:bandstruct}(c)] reveals that they originate from a
region with a parabolic rather than a linear $E(k)$ dispersion. At
these energies, the 2D bandstructure $E(k_x, k_y)$
[Fig.~\ref{fig:3D_weak}] near the satellites does not show cone-like
structures unlike near the main Dirac cone. Consequently, the Landau
levels show a Schr\"odinger-like rather than a Dirac-like
[Eq.~(\ref{eq:landaulevels})] B-field dependence as a function of
$E_F$. However, when plotted as a function of the charge carrier
number $N$ or, equivalently, as a function of $V_{\mathrm{BG}}$ a
linear $B-N$ ($B-V_{\mathrm{BG}}$) dependence emerges
[Fig.~\ref{fig:DOS_weak}(b)] for both the main Dirac cone and the
satellites and this discriminating feature is lost. Such a linear
dependence was seen in the experiment for the
satellites,\cite{Gorbachev14, Hunt13} and was attributed to a Dirac
cone replica. It is the absence of the 0$^{th}$ Landau level and its
Hofstadter butterfly for the satellites both in experiment and in our
simulation that unambiguously confirms that the satellites are
associated with a parabolic band structure rather than with a Dirac
cone. We note that, strictly speaking, no real cone structures appear
in the 2D bandstructure at the energy of the satellite states for the
present moir\'e potential. All cone-like dispersions visible in a 1D
cut at $k_y=0$ show, upon consideration of the full 2D bandstructure,
only avoided crossings [Fig.~\ref{fig:3D_weak}(b)], and not a true
Dirac cone [Fig.~\ref{fig:3D_weak}(a)].

The origin of the non-relativistic dispersion can be easily traced to
the unperturbed spectrum of bulk graphene
[Fig.~\ref{fig:bandstruct}(a)]. The satellites emerge from parabolic
bands with energies $\left|\Delta E\right| \approx 0.15$ eV above and
below the Dirac points. Replica D$_1$ of the Dirac cone centered at
$k_y=0$ appear at much higher energies $\left|\Delta E\right| \gtrsim
0.28$ eV and are submerged in a region of high DOS. Accordingly, the
superlattice potential resulting from the van der Waals interaction
with the hBN substrate, which is of the order $\left|V\right|\lesssim
0.1$ eV, represents only a moderately weak perturbation of the
parabolic bands giving rise to distortion and narrow avoided crossings
but cannot significantly shift the distant Dirac cone into the region
of low DOS and into the proximity of D$_0$. In turn, increasing the
van der Waals interaction to an unrealistic strength with on-site
potential $V_V=-0.5$ eV [Fig.~\ref{fig:bandstruct}(e)] renders the
replica D$_1$ of the Dirac cone visible near $E=0.114$ eV (see
Fig.~\ref{fig:3D_strong}) in addition to the main cone D$_0$ at
$E=-0.07$ eV. We note that the main cone D$_0$ shows a finite gap of
about $10$ meV, while the replica cone appears gapless [compare
  Fig.~\ref{fig:3D_strong} (a) and (b)]. The DOS in the $B$-$E$
representation [Fig.~\ref{fig:DOS_strong}(a)] and $B$-$N$
representation [Fig.~\ref{fig:DOS_strong}(b)] show now two emerging
fans of relativistic dispersion $B\sim E^2$
[Fig.~\ref{fig:DOS_strong}(a)] and linear dispersion $B\sim N$. The
Fermi velocity of the main cone is $v_{F}=0.62 \cdot 10^6$ m/s, while
we find $v_{F}=0.39 \cdot 10^6$ m/s for the secondary cone. The
calculated density of states [see Fig.~\ref{fig:DOS_strong}$\,$] shows
the presence of Landau levels of Dirac fermions emerging from two
Dirac points. The curvature of the parabolas determined by
$v_{\mathrm{F}}$ again fits well to the DOS near both D$_0$ and D$_1$
[see red dashed parabolas in Fig.~\ref{fig:DOS_strong}(a)]. The
additional features emerging near $E=0.03$ eV and $E=-0.19$ eV at
$B=0$ [see arrows in Fig.~\ref{fig:DOS_strong}(a)] resulting from the
regions of non-relativistic quasi-quadratic dispersion remain present
for this much stronger superlattice potential. Thus, the coexistence
in the spectrum of a Dirac-like and a Schr\"odinger-like diamagnetic
response persists. As a function of back gate voltage, the satellite
structures for relativistic and non-relativistic particles
[Fig.~\ref{fig:DOS_strong}(b)] are similar. However, as discussed
above, they can be well distinguished by the presence of the zeroth
Landau level for relativistic dispersion. A more direct approach to
observe the two different dispersion relations would be a direct
energy dependent measurement.  For example, measuring the optical
transitions within the satellite structures and within the Landau
levels of the main cone\cite{MagnetoOptic} would allow to distinguish
between these dispersion relations.

While we do not specifically address disorder in this work, we surmise
the satellite structures induced by the moir\'e pattern will respond
drastically different to different classes of disorder. Short-range
disorder such as lattice vacancies softens all features of the DOS, as
numerical tests have shown. However, at realistic disorder
concentration the Landau levels and satellites are still well
discernable in line with previous studies.\cite{Libisch10} Long-range
disorder, on the other hand, may strongly wash out these structures.

\section{Conclusion}
\label{sec:conclusion}
We have simulated the electronic structure of a large graphene flake
on a hexagonal boron nitride substrate as a function of a
perpendicular magnetic field. We have shown that the periodic moir\'e
potential leads to the formation of the Hofstadter butterfly and
satellite structures. For a realistic substrate potential, satellites
close to the Dirac point feature a parabolic dispersion and cannot be
considered replica of the Dirac cone. We have observed that when the
$B$ field dependence is measured as a function of the back gate
voltage rather than the energy Landau levels for Schr\"odinger-like
and Dirac-like dispersion display the same (approximately) linear
behavior. As the unambiguous hallmark for the (non) relativistic
response we have identified the (absence) presence of the zeroth
Landau level at the satellite and its associated Hofstadter butterfly
structure. Our findings thus suggest that the absence of Hofstadter
butterfly structure in recent experiments is due to the fact that the
observed satellite structure results from parabolic bands rather than
replica of the Dirac cone.

\section*{Acknowledgments}
\label{sec:acknowledgments}
We gratefully acknowledge support from the doctoral colleges CMS (TU
Vienna) and Solids4Fun (FWF), as well as by ViCom (SFB 041-ViCom).

\end{document}